\documentclass[%
 aip,
 amsmath,amssymb,
 reprint,%
]{revtex4-1}

\usepackage{graphicx}
\usepackage{dcolumn}
\usepackage{bm}

\usepackage[utf8]{inputenc}
\usepackage[T1]{fontenc}
\usepackage{mathptmx}
\usepackage{etoolbox}
\usepackage{siunitx}
\DeclareUnicodeCharacter{0301}{\'{e}}

\makeatletter
\def\@email#1#2{%
 \endgroup
 \patchcmd{\titleblock@produce}
  {\frontmatter@RRAPformat}
  {\frontmatter@RRAPformat{\produce@RRAP{*#1\href{mailto:#2}{#2}}}\frontmatter@RRAPformat}
  {}{}
}%
\makeatother
\begin{document}

\preprint{AIP/123-QED}

\title[]{Polariton Enhanced Free Charge Carrier Generation in Donor–Acceptor Cavity Systems by a Second-Hybridization Mechanism}
\author{Weijun Wu}

\author{Andrew E. Sifain}

\author{Courtney A. Delpo}

\author{Gregory D. Scholes}
 \email{gscholes@princeton.edu}
\affiliation{%
Department of Chemistry, Princeton University, Princeton, NJ 08540, U.S.
}%

\date{\today}

\begin{abstract}
Cavity quantum electrodynamics has been studied as a potential approach to modify free charge carrier generation in donor-acceptor heterojunctions because of the delocalization and controllable energy level properties of hybridized light–matter states known as polaritons. However, in many experimental systems, cavity coupling decreases charge separation. Here, we theoretically study the quantum dynamics of a coherent and dissipative donor–acceptor cavity system, to investigate the dynamical mechanism and further discover the conditions under which polaritons may enhance free charge carrier generation. We use open quantum system methods based on single-pulse pumping to find that polaritons have the potential to connect excitonic states and charge separated states, further enhances free charge generation on an ultrafast timescale of several hundred femtoseconds. The mechanism involves that polaritons with proper energy level allow the exciton to overcome the high Coulomb barrier induced by electron-hole attraction. Moreover, we propose that a second-hybridization between a polariton state and dark states with similar energy enables the formation of the hybrid charge separated states that are optically active. These two mechanisms lead to a maximum of 50\% enhancement of free charge carrier generation on a short timescale. However, our simulation reveals that on the longer timescale of picoseconds, internal conversion and cavity loss dominate and suppress free charge carrier generation, reproducing the experimental results. Thus, our work shows that polaritons can affect the charge separation mechanism and promote free charge carrier generation efficiency, but predominantly on a short timescale after photoexcitation.
\end{abstract}

\maketitle

\section{\label{sec:level1}Introduction}

The excitonic character of organic semiconductors is crucial for designing the architecture and geometry of organic photovoltaic cells. A critical component is the donor-acceptor heterojunction, where exciton goes dissociation and charge transfer.\cite{bredas2009molecular,laird2007advances} The optimization of donor-acceptor materials' efficiency for free charge carrier generation is based on the control of the basic ultrafast dynamics processes including optical absorption and exciton formation \cite{kim2007efficient, bredas1985polarons}, exciton transport \cite{ahn2008experimental, silinsh1997organic, yoo2004efficient}, exciton dissociation and charge separation\cite{morteani2004exciton, hwang2007ultrafast,drori2008below} and free charge-carrier mobility \cite{coropceanu2009interaction}. However, the mechanism at the molecular level and the coherent quantum-mechanical description of the dynamics are still topics of debate and investigation. 

\begin{figure}
    \includegraphics[width=0.45\textwidth]{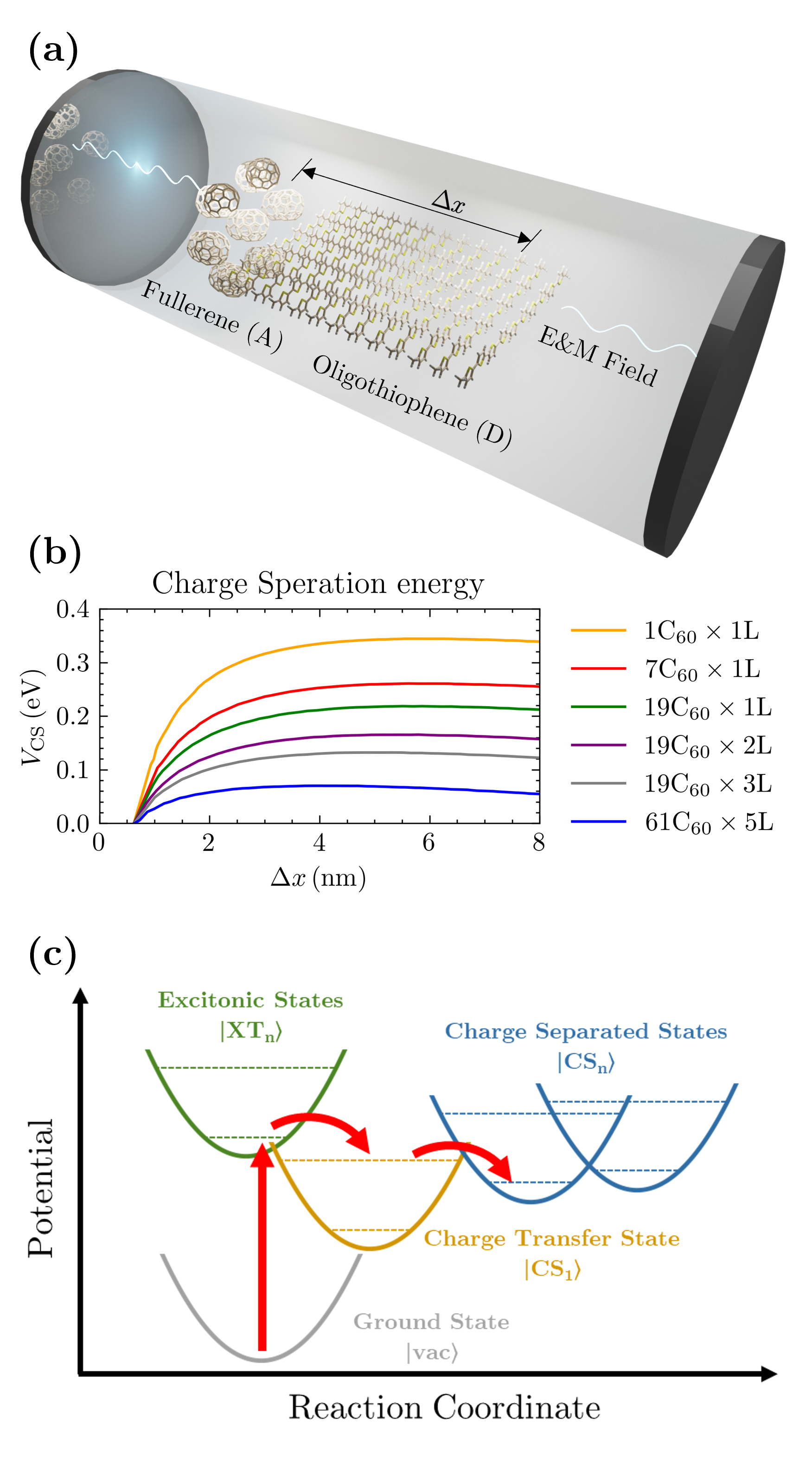}
    \caption{\label{fig:polymer}(a) An illustration of the oligothiophene:fullerene heterojunction embedded in a Fabry–Pérot cavity. The fullerenes of 2 hexagon layers with 7 fullerenes per layer (7C$_{60}\times2$L) is displayed. The fullerenes are treated as one super-molecule and the distance between fullerene super-molecule and a specfic layer of oligothiophene polymer chain is labeled as $\Delta x$. (b) The Coulomb attraction potential between a hole on oligothiophene polymer chain and an electron on fullerene super-molecule with a distance of $\Delta x$ for different packing types for fullerenes ($z$ layers with $n$ fullerenes each layer, $z$C$_{60}\times n$L in short). The data comes from Ref.~\onlinecite{tamura2013ultrafast,tamura2013potential} by Density Functional Theory calculation. (c) Concept of charge separation mediated by vibronic charge transfer states as explained in Ref.~\onlinecite{tamura2013ultrafast}. The dashed lines of each potential curves represent the vibrational states. Red arrows represent population transfer.}
\end{figure}

It has already been theoretically proposed that the electron-hole separation at the interface is significantly influenced by the contradictory relation between the Coulomb attraction barrier and vibronic charge transfer state.\cite{tamura2013ultrafast} First, increasing delocalization of charges can substantially reduce the Coulomb barrier\cite{tamura2013potential} (FIG.~\ref {fig:polymer}(b)). Second, the excess energy of the exciton-charge transfer transition can induce vibrational excitations. The vibronic charge transfer state can be resonant with charge separated states with high potential and thus overcome the Coulomb barrier\cite{marcus1993electron, dimitrov2012energetic, grancini2013hot} (FIG.~\ref {fig:polymer}(c)). It has also been indicated that internal conversion and non-radiative transition could modify the charge separation dynamics.\cite{huix2015concurrent} 

Cavity quantum electrodynamics (cavity QED)\cite{raimond2001manipulating,garcia2021manipulating} has been studied recently to modify the excitonic character through light-dress, via embedding the materials between two reflective mirrors (known as the Fabry–Pérot cavity). Strong coupling between the polarization of a material and a quantized electromagnetic field confined in a cavity forms hybridized light-matter states known as polaritons. Polaritons have extraordinary delocalization in the molecule basis and modified energy levels. The collective nature of polaritons due to the coupling between many molecules and a single global oscillator (i.e., photon) is central to extraordinary delocalization phenomena including long-range energy transfer, \cite{zhong2017energy,du2018theory,rozenman2018long} enhanced charge conductivity\cite{orgiu2015conductivity,hagenmuller2017cavity,nagarajan2020conductivity} and superconductivity \cite{schlawin2019cavity}. The control and change of chemistry at select energies has been a central topic including modification of ground\cite{thomas2016ground,lather2019cavity,thomas2019tilting,li2021theory} and excited state chemical reactions,\cite{hutchison2012modifying,galego2016suppressing,mandal2019investigating,mony2021photoisomerization}, singlet fission\cite{climent2022not,martinez2018polariton} and spin state selectivity.\cite{stranius2018selective,polak2020manipulating,yu2021barrier,ye2021direct} 

The delocalization nature of the polaritons is predicted to improve exciton migration and charge transport.\cite{garcia2021manipulating}Studies on charge transfer in cavities\cite{mandal2020polariton} also show the potential of charge transfer enhancement in a dimer system in the Marcus Inverted Regime \cite{semenov2019electron} or under incoherent driving conditions\cite{wellnitz2021quantum}. However, at the interface, there is a competition between charge transfer from the polariton and the decay of the polariton to the ground state. A recent spectroscopic study on P3HT(poly(3-hexylthiophene)):PCBM(phenyl-C61-butyric acid methyl ester) inside the cavity suggests that although the rate of charge transfer from the lower polariton is slowed relative to the rate of charge transfer from a bare P3HT polymers, charge transfer from polariton to create free charge carriers remains fast enough to compete with the decay of the polariton to the ground state.\cite{delpo2021polariton} 

In order to understand the competition between polariton mediated free charge carrier generation and the dynamics of dissipation and dephasing, we theoretically utilize a general open quantum system method to study the coherent and dissipative quantum dynamics in both bare and cavity donor-acceptor systems. We elucidate the roles that polaritons may play on the charge separation mechanism, analyze the dynamical reason for the experimental negative effect of polaritons, and furthermore, clarify the conditions of certain time-scale and coupling strength under which the polaritons may positively enhance free charge carriers generation.

Here, we propose that adjusting the energy level of the polaritons may change the relation between the height of the Coulomb barrier and the energy of vibronic charge transfer state to promote exciton dissociation at the interface.\cite{tamura2013ultrafast, tamura2013potential, huix2015concurrent} Moreover, we propose a novel mechanism that a second-hybridization (distinct from the light-matter hybridization that forms polaritons) between polaritons and dark states can create the polariton states with charge separation components working as instantaneous charge generators on a short timescale after photoexcitation. Our quantum dynamics simulation with Redfield Theory for vibronic coupling and Lindblad Theory for cavity loss working on a phenomenological Hamiltonian spanned by Frenkel exciton states and charge separated states predicts that single-pulse-pumping polaritons can significantly enhance free charge carrier generation up to 50\% comparing to the bare system on a short timescale of several hundred femtoseconds, before internal conversion and cavity loss dominates.

\section{\label{sec:level2}Hamiltonian and Dynamics Model}
We start by specifying a basis set of the Hilbert Space to build the Hamiltonian as an extension of the model in Ref.~\onlinecite{huix2015concurrent}. To provide realistic parameters representing regioregular polymer:fullerene heterojunctions, we consider the parent compounds, oligothiophene:fullerene in FIG.~\ref {fig:polymer}(a), as a model system. The fullerene acceptor is represented by an effective, coarse-grained super-molecule, labeled as $\mathrm{A}_0$, while the primary oligothiophene donor photoexcitations in the self-assembled lamellae structures are interchain, H-aggregate type excitons. \cite{jiang2002spectroscopic} Intrachain interactions are eliminated by Fourier Transformation (see Supplementary Material S1.1). Therefore, $N$ oligothiophene chains are treated explicitly as individual sites labeled as $\mathrm{D}_{i}$, where $i=1,2,...,N$ represents the $i$-th site from fullerenes. Furthermore, we define the localized Frenkel exciton configurations on oligothiophene donor as $|\mathrm{XT}_i\rangle =|\mathrm{D}_{i}^{-}\rangle \otimes |\mathrm{D}_{i}^{+}\rangle $ and the non-localized electron-hole pair with the one hole on oligothiophene and one electron on fullerenes as $|\mathrm{CS}_i\rangle =|\mathrm{A}_{0}^{-}\rangle \otimes |\mathrm{D}_{i}^{+}\rangle $. Since the distinction between charge transfer state and charge separated state is not rigorous, the charge transfer state is roughly defined as $|\mathrm{CS}_{i=1}\rangle$. The quantized electromagnetic field confined in the cavity with single mode approximation are written in the Fock Space. For example, $|1\rangle$ is the single photon state. A common ground state without any excitation, i.e. vacuum state $|\mathrm{vac}\rangle $, is of zero energy and can be a sink for dissipation processes.

A phenomenological Hamiltonian with $N=13$ oligothiophene sites can be represented within the basis set defined above after single excitation truncation
\begin{subequations}
\begin{align}
H_{\mathrm{e}}
={}&
\epsilon ^{\mathrm{XT}}\sum_i{|\mathrm{XT}_i\rangle \langle \mathrm{XT}_i|}+J_{\mathrm{XT}}\sum_i{\left( |\mathrm{XT}_i\rangle \langle \mathrm{XT}_{i+1}|+\mathrm{h}.\mathrm{c}. \right)}
\label{Hamiltoniana}{}\\&
+\sum_i{\epsilon _{i}^{\mathrm{CS}}|\mathrm{CS}_i\rangle \langle \mathrm{CS}_i|}+J_{\mathrm{CS}}\sum_i{\left( |\mathrm{CS}_i\rangle \langle \mathrm{CS}_{i+1}|+\mathrm{h}.\mathrm{c}. \right)}
\label{Hamiltonianb}{}\\&
+J_{\mathrm{int}}\left( |\mathrm{XT}_1\rangle \langle \mathrm{CS}_1|+\mathrm{h}.\mathrm{c}. \right) 
\label{Hamiltonianc}{}\\&
+\hbar \omega _c|1\rangle \langle 1|+g\sum_i{\left( |1\rangle \langle \mathrm{XT}_i|+\mathrm{h}.\mathrm{c}. \right)}
\label{Hamiltoniand}
\end{align}
\label{equ:Hamiltonian}
\end{subequations}

The Frenkel Exciton states $|\mathrm{XT}_i\rangle$ are assumed to be spatially invariant with the identical on-site energies $\epsilon ^{\mathrm{XT}}=\SI{2.3}{\eV}$  as a typical value. The relative energies of the charge separated states $\epsilon _{i}^{\mathrm{CS}} - \epsilon _{1}^{\mathrm{CS}}$ are defined by the Coulomb barrier in FIG.~\ref{fig:polymer}(b) with the inter-oligothiophene distance of \SI{0.38}{\nm}\cite{tamura2013ultrafast,tamura2013potential}. The potential energy surface for CS state is approximately unaffected by dipole self energy \cite{mandal2020polariton} since the permanent dipole of CS states should be parallel to the standing wave vector, as shown in FIG.~\ref{fig:polymer}(a). The excess energy of the interfacial exciton-charge transfer transition that induces a vibronic charge transfer state is expressed by the energetic offset between the interfacial excitonic and charge transfer states $\Delta E_{\mathrm{offset}}=\epsilon ^{\mathrm{XT}}-\epsilon _{1}^{\mathrm{CS}}$. The other intermolecular coupling parameters are determined from ab-initio theory calculation by Tamura and coworkers\cite{tamura2013ultrafast,tamura2013potential,huix2015concurrent,polkehn2018quantum,tamura2011exciton}: The nearest-neighbour coupling within the H-aggregate excitonic states and the charge separated states are $J_{\mathrm{XT}} = \SI{0.1}{\eV}$ and $J_{\mathrm{CS}} = \SI{-0.12}{\eV}$, respectively. Charge transfer process at the interface can be described by the strength $J_{\mathrm{int}} = \SI{0.2}{\eV}$. 

The cavity photon mode frequency $\omega_{\mathrm{c}}$ and collective light-matter strong coupling strength $G=g\sqrt{N}$ under the rotating wave approximation are adjustable by cavity preparation.\cite{ebbesen2016hybrid} Experimentally, one can change the inter-mirror distance to modify $\omega_{\mathrm{c}}$ as the frequency of a standing wave, and change the number of molecules in one oligothiophene chain $N_y$ to modify single site-photon coupling strength because $g\propto \sqrt{N_y}$. $N_y$ can be large enough to have the system reach the strong light-matter coupling regime in a typical Fabry–Pérot micro-cavity (2 mirrors in the size of 1 inch by 1 inch and the distance between them of several hundred nanometers). In this model, we assume cavity photon is single mode as a standing wave under long-wavelength limit, while other work reports the cases where cavity photons have non-zero in-mirror-plan momentum.\cite{tichauer2021multi} Only XT states can effectively achieve light-matter coupling, while CS are decoupled from the cavity, since CS states have negligible transition strength, regardless of the cavity, because the transition dipole depends on electron-hole overlap.

Vibronic coupling induced internal conversion and dephasing for the electronic density matrix $\rho_{\mathrm{e}}\left( t \right)$ is calculated by Bloch-Redfield Theory. Due to the weak electronic-vibrational coupling $H_\mathrm{I}$ \cite{huix2015concurrent}, Quantum Markovianity can deal with the vibrational modes as a large thermal reservoir at thermal-equilibrium $\rho _{\mathrm{v}}$ \cite{jeske2015bloch}. The quantum master equation in the Dirac Picture (represented by tildes) reads
\begin{equation}
\frac{d\tilde{\rho}_{\mathrm{e}}\left( t \right)}{dt}=-\frac{1}{\hbar ^2}\int_0^{\infty}{d\tau}\mathrm{Tr}_{\mathrm{v}}\left[ \tilde{H}_I\left( t \right) ,\left[ \tilde{H}_I\left( t-\tau \right) ,\tilde{\rho}_{\mathrm{e}}\left( t \right) \otimes \rho _{\mathrm{v}} \right] \right] 
\label{equ:Redfield}
\end{equation}
Lindbladians $\gamma _X\mathcal{L} _X\left[ \rho _{\mathrm{e}} \right] =\gamma _X\left( X\rho _{\mathrm{e}}X^{\dagger}-\frac{1}{2}\left\{ X^{\dagger}X,\rho _{\mathrm{e}} \right\} \right)$ are added to Eq.~(\ref{equ:Redfield}) for cavity loss ($X=|\mathrm{vac}\rangle \langle 1|$, $\gamma _X=\kappa=20\mathrm{THz}$) and electron-hole recombination of Frenkel exciton ($X=|\mathrm{vac}\rangle \langle \mathrm{XT}_i|$, $\gamma _X=\gamma_\mathrm{decay}=1\mathrm{GHz}$). The free charge carrier is described by the charge separation beyond $|\mathrm{CS}_N\rangle $ and its generation is expressed as a Lindbladian with $X=|\mathrm{vac}\rangle \langle \mathrm{CS}_N|$ and $\gamma _X = \gamma _\mathrm{out} = 10\mathrm{THz}$. The observable of interest, the free charge carrier generation rate, can be obtained by
\begin{equation}
I_{\mathrm{FC}}(t)=-\gamma _{\mathrm{out}}\mathrm{Tr}\left( P_{\mathrm{excitation}}\mathcal{L}_{|\mathrm{vac}\rangle \langle \mathrm{CS}_N|}\left[ \rho _{\mathrm{e}}\left( t \right) \right] \right) 
\label{equ:FCCurrent}
\end{equation}
where $P_{\mathrm{excitation}}=\sum_i{|\mathrm{XT}_i\rangle \langle \mathrm{XT}_i|}+\sum_i{|\mathrm{CS}_i\rangle \langle \mathrm{CS}_i|}+|1\rangle \langle 1|$ is the projector operator to the single excitation subspace. The details of the theory are in Supplementary Material S1.3. 

This Hamiltonian and quantum dynamics framework provides a lens into the complex interplay between light-matter coupling, delocalization, thermalization, dephasing and dissipation.

\section{\label{sec:level3}results and discussion}
In section~\ref{sec:level3A} and \ref{sec:level3B}, we mainly focus on the case of fullerene packing type 7C$_{60}\times1$L (Coulomb barrier \SI{0.25}{\eV}), interfacial energy offset $\Delta E_{\mathrm{offset}}=\SI{0}{\eV}$ (no initial vibronic effect), photon mode $\omega_\mathrm{c}= \SI{2.495}{\eV}$ and collective light-matter coupling $G=\SI{0.12}{\eV}$, as an example to illustrate the effect of polaritons. The cases for different parameters will be discussed in section~\ref{sec:level3C}.

\begin{figure*}
    \includegraphics[width=1\textwidth]{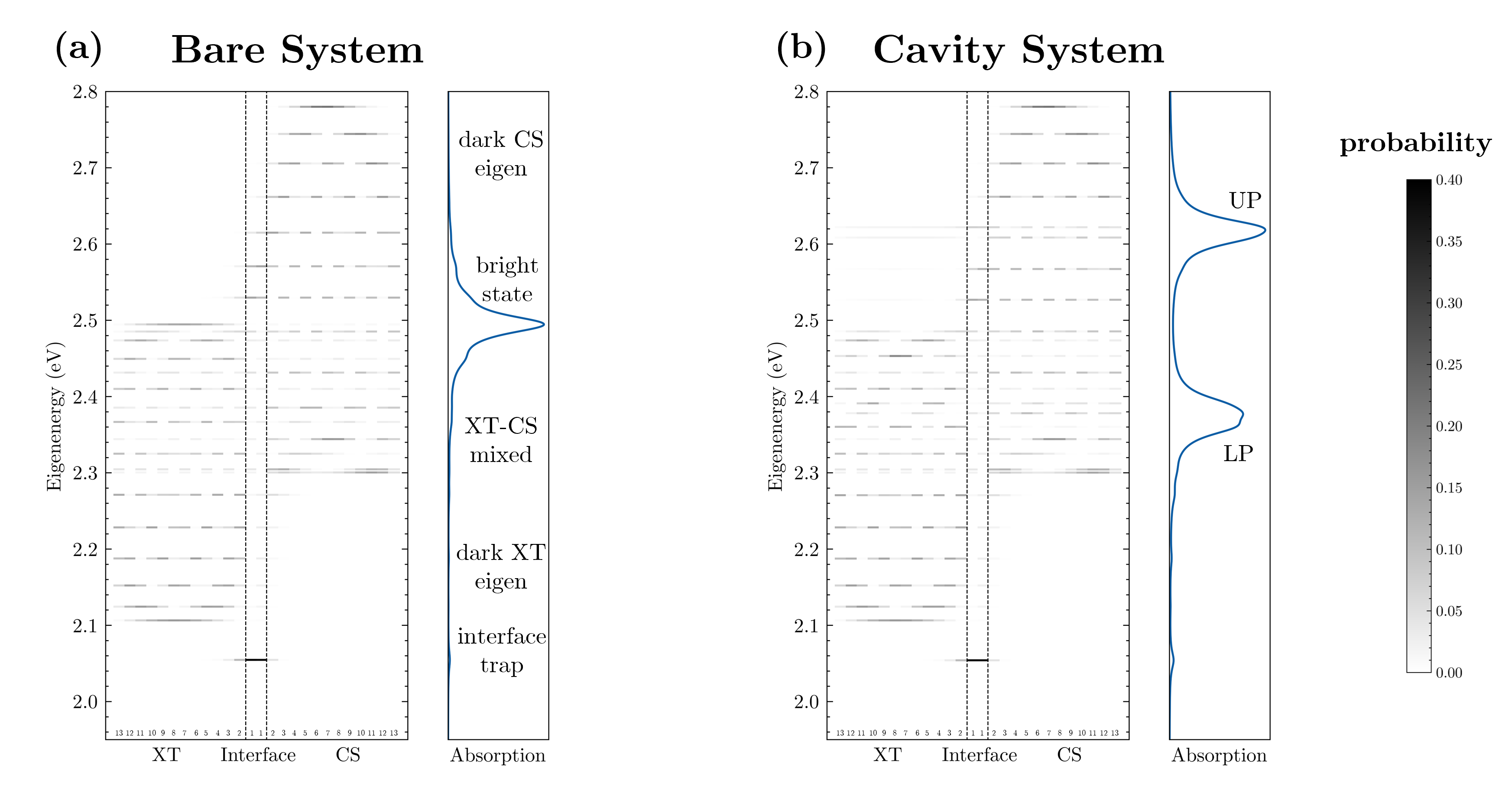}
    \caption{Electronic excited eigenstate of the system with fullerene packing type 7C$_{60}\times1$L and $\Delta E_{\mathrm{offset}}=\SI{0}{\eV}$ for (a) Bare System and (b) Cavity System ($\omega_\mathrm{c}=\SI{2.495}{\eV}$ and $G=\SI{0.12}{\eV}$). In the left panel, the ordinate defines the eigenvalues, while the abscissa defines a series of basis states pertaining to the XT manifolds ($|\mathrm{XT}_{N}\rangle, ...\ , |\mathrm{XT}_{2}\rangle$ from left to right), the subset of interfacial states ($|\mathrm{XT}_{1}\rangle$ and $|\mathrm{CS}_{1}\rangle$) and the CS manifolds ($|\mathrm{CS}_{2}\rangle, ...\ , |\mathrm{CS}_{N}\rangle$ from left to right). Photon is not displayed. The probabilities are represented as a density profile. The right panel is the absorption spectrum fitted from the vacuum-to-eigen transition dipole. Bright state, interface trap, upper polariton (UP) and lower polariton (LP) are marked.}
    \label{fig:eigen}
\end{figure*}

\subsection{\label{sec:level3A}Eigenspectrum Calculation}
FIG.~\ref{fig:eigen}(a) depicts the eigenstates and absorption spectrum of the bare system. A low-energy localized interface trap state emerges, where an energy gap of \SI{0.25}{\eV} between the interface trap and the lowest delocalized CS manifolds corresponds to the Coulomb barrier for fullerene packing type 7C$_{60}\times1$L. At higher energies, the delocalized excitonic manifolds become mixed with delocalized charge separated states below the effective top of the barrier, suggesting a tunneling effect, which is beneficial for free charge carrier generation. However, the dark nature of these XT-CS mixed state prevents populating from photoexcitation but internal conversion only. The absorption spectrum demonstrates a bright state of the H-aggregate at the upper edge of the XT manifolds, mixing weakly with the CS manifolds. Above the bright state, there are some delocalized and dark CS eigenstates. Thus by pumping the bright state, free charge carrier generation mostly occurs after population transfer downhill to dark XT-CS mixed state or uphill to dark CS eigenstates.

FIG.~\ref{fig:eigen}(b) depicts the eigenstates and absorption spectrum of the cavity system, where the photon mode $\omega_\mathrm{c}=\SI{2.495}{\eV}$ is resonant with the bare bright state to achieve strong light-matter hybridization. The upper polariton ($E_\mathrm{UP} = \SI{2.61}{\eV}$) and lower polariton ($E_\mathrm{LP} = \SI{2.36}{\eV}$) show up with larger broadening than the bare bright state, signifying inhomogeneous broadening (see Supplementary Material S2.2).

Within the absorption linewidth, the polaritons present large CS components. As shown in FIG.~\ref{fig:eigen}, the CS fraction for upper and lower polariton branchs are 51\% and 34\% respectively, while the CS fraction for the bare bright state is only 7\%. In contrast to the bare bright state, the optically active polaritons are direct and instantaneous free charge carrier generators.

The inhomogeneous broadening and CS components features indicate that polaritons $|\mathrm{UP\left(+\right)/LP\left(-\right)}\rangle$ are not merely a hybridization between photon $|1\rangle$ and bare bright state $|\mathrm{B}\rangle$, but also involve mixing dark states $|d\rangle$, which could be understood by first-order perturbation theory (see Supplementary Material S1.4)
\begin{equation}
|\mathrm{UP\left(+\right)/LP\left(-\right)}\rangle = \frac{1}{\sqrt{2}}\left( |1\rangle \pm |\mathrm{B}\rangle \right) +\sum_d{\frac{{{g_{d}^{'}}/{\sqrt{2}}}}{E_\mathrm{B} \pm \left| g_{B}^{'} \right|-E_d}|d\rangle}
\end{equation}
where $g_{\mathrm{B}}^{'}=\vec{\varepsilon}\cdot \langle \mathrm{B}|\vec{\mu}|\mathrm{vac}\rangle \approx G$ and $g_{d}^{'}=\vec{\varepsilon}\cdot \langle d|\vec{\mu}|\mathrm{vac}\rangle $ are the field $\vec{\varepsilon}$ - dipole $\vec{\mu}$  interaction for bright and dark state respectively, while $E_\mathrm{B}$ and $E_d$ are the unperturbed energy of bright and dark state respectively. Second-hybridization occurs when unperturbed polaritons have similar energy to certain unperturbed dark states with non-zero transition dipoles (despite the transition dipoles are extremely tiny). The resulting perturbed polaritons contain significant CS component, which is beneficial for free charge carrier generation.

\begin{figure*}
    \includegraphics[width=0.9\textwidth]{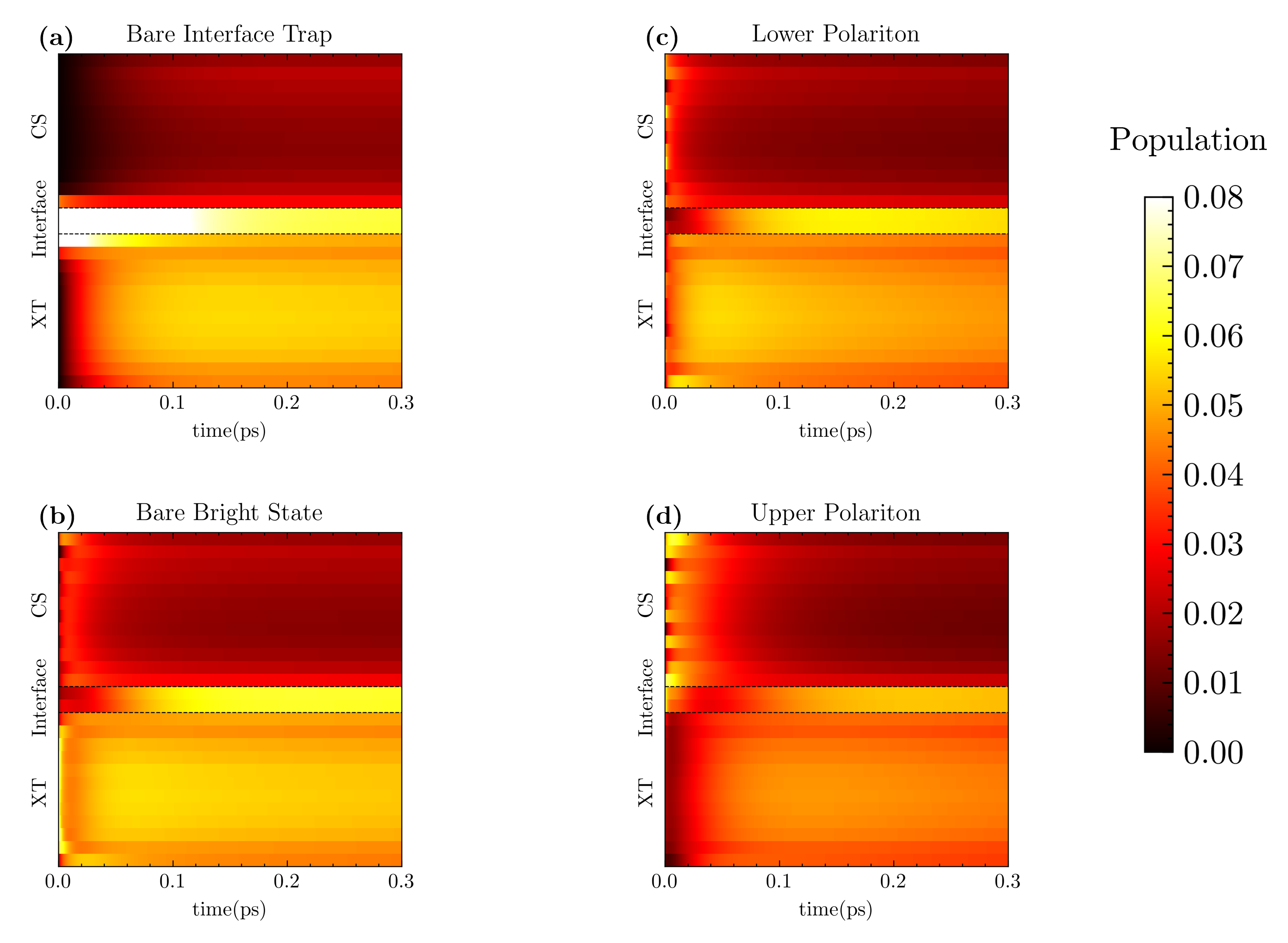}
    \caption{Population dynamics from quantum dynamics simulation of fullerene packing type 7C$_{60}\times1$L and $\Delta E_{\mathrm{offset}}=\SI{0}{\eV}$ with different initial condition: for bare system (a) interface trap state, (b) bright state and for cavity system ($\omega_\mathrm{c}=\SI{2.495}{\eV}$ and $G=\SI{0.12}{\eV}$) (c) lower polariton, (d) upper polariton. The abscissa defines the time, while the ordinate, similar to FIG.~\ref{fig:eigen}, defines a series of basis states ($|\mathrm{XT}_{N}\rangle, ...\ , |\mathrm{XT}_{1}\rangle$, $|\mathrm{CS}_{1}\rangle, ...\ , |\mathrm{CS}_{N}\rangle$ from bottom to top). Photon is not displayed. The population is represented as a density profile.}
    \label{fig:dynamics}
\end{figure*}

\subsection{\label{sec:level3B}Dynamics Simulation}

\begin{figure}
    \includegraphics[width=0.45\textwidth]{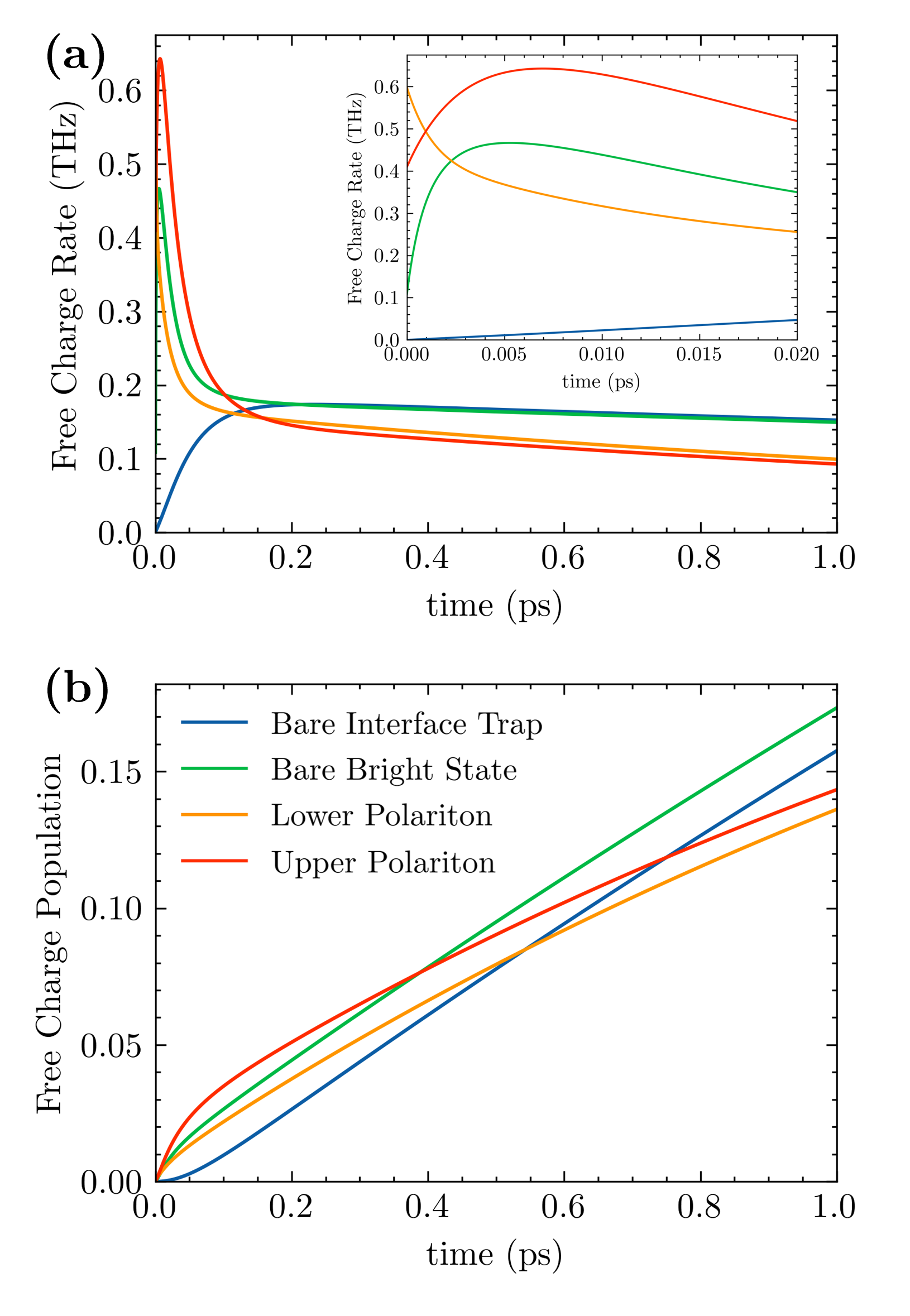}
    \caption{The extracted free charge carrier information of the four initial conditions from quantum dynamics simulation for FIG.~\ref{fig:dynamics}:  (a) Free charge carrier generation rate at each time step from Eq.~(\ref{equ:FCCurrent}) up to \SI{1}{\ps}. The inset is a zoomed-in area of the first \SI{0.02}{\ps}. (b) Free charge carrier population dynamics by integration for the corresponding rates in (a).}
    \label{fig:FreeCharge}
\end{figure}

Based on the eigenstates calculation (FIG.~\ref{fig:eigen}), Redfield and Lindblad Theory simulation for different cases are compared in FIG.~\ref{fig:dynamics}, while rates and population calculated by Eq.~(\ref{equ:FCCurrent}) for free charge generation are shown in FIG.~\ref{fig:FreeCharge}.

When the bare system is initially populated at the interface trap (FIG.~\ref{fig:dynamics}(a)), the population transfers to the delocalized XT manifolds within \SI{0.01}{\ps} before transfering to the CS manifolds in \SI{0.04}{\ps}. If the bare bright state is pumped initially (FIG.~\ref{fig:dynamics}(b)), the population of delocalized bright exciton partially transfers to the CS manifolds. However, on a longer timescale of \SI{0.3}{\ps}, most of the population still gets trapped at the interface and the XT population overwhelms the CS population.

On the contrary, when the polaritons are pumped for the cavity system, second-hybridization between polaritons and dark states populates the CS manifolds immediately. For the lower polariton case (FIG.~\ref{fig:dynamics}(c)), XT manifolds population rises rapidly, while the CS manifolds population will transfer to interface trap after \SI{0.02}{\ps}, corresponding to the population transfer from lower polariton to XT-CS mixed state. For the upper polariton case (FIG.~\ref{fig:dynamics}(d)) there is significant population at the CS manifolds until \SI{0.06}{\ps} due to the downhill transfer to dark CS eigenstates before deexcitation to XT manifolds. XT manifolds population of upper polariton case is less than that for lower polariton cases. In all the four cases, the systems seem to reach a quasi-steady state and the CS manifolds populate more around $\mathrm{CS}_1$ and $\mathrm{CS}_N$.

The free charge carrier generation for the four cases in FIG.~\ref{fig:FreeCharge} better demonstrates the effect of cavity mediated charge separation. In the first \SI{0.1}{\ps} of FIG.~\ref{fig:FreeCharge}(a), the rates of the four cases are sorted in the order of the energy of the states, meaning that the upper polariton can overcome the Coulomb barrier more easily than the bare bright state by internal conversion to the CS manifolds to maintain a larger CS population. This is similar to effect demonstrated in Ref.~\citenum{huix2015concurrent} that the excess energy of interfacial XT-CS transfer can induce charge transfer states with high vibrational level to pass the Coulomb barrier. Internal conversion induced population transfer from CS manifolds to interface and XT manifolds decreases the rates dramatically for the polaritons and bare bright state cases,  while for the bare interface trap case the rate increases. 

After \SI{0.2}{\ps}, all the four cases have reached a quasi-steady state where the internal conversion is much faster than any types of dissipation (including cavity loss, electron-hole recombination and and free charge carrier generation), resulting in a proportional depopulation among all the eigenstates. Thus, the rates for both polariton cases converge to a linear decrease on a long timescale. The two bare system cases show the same feature.

The emergence of the quasi-steady states implies that the interesting dynamics mainly occur on the an ultrafast timescale of 100fs, within which the hybridized polaritons are more beneficial for free charge carrier generation than bare bright state. A more detailed free charge carrier dynamics in a short timescale is displayed in the inset of FIG.~\ref{fig:FreeCharge}(a) up to 20fs. At the first 1fs, both polariton cases show much larger rates than the bare system cases, meaning that the polaritons with large CS component can achieve instantaneous charge generators, due to the second-hybridization. More surprisingly, the lower polariton shows a larger rate than that of upper polariton due to larger $\mathrm{CS}_N$ component. FIG.~\ref{fig:FreeCharge}(b) shows the free charge carrier population by integrating the rates, where the advantage of the upper polariton can be held up to \SI{0.4}{\ps}.

However, after \SI{0.4}{\ps}, the population for the bare bright state case exceed the population for the upper polariton case. Because on a long timescale, the converged free charge generation rate for the bare system is higher than that for the cavity system, consistent with the experimental results \cite{delpo2021polariton}. This can be explained by the following two reasons. First, the quasi-steady states are not as beneficial as initial polaritons for free charge carrier generation, because the large CS components feature for polaritons has been eroded by internal conversion. Thus, second-hybridized polaritons can only work as "instantaneous" charge generators after photoexcitation. Second, the relatively short lifetime of cavity photon ($\kappa^{-1}=50\mathrm{fs}$) induces an additional dissipation channel. The total effect of population decay by cavity loss can be found in Supplementary Material S2.1. By preparing a high Q-factor cavity, for example a pair of thicker mirrors, the cavity photon lifetime can increase. In Supplementary Material S2.1 we demonstrate an ideal simulation for the free charge carrier generation rates of a lossless cavity system, where the cavity system and bare system converge to the same quasi-steady state and upper polariton works better than bare bright state even in the long timescale. However, the detrimental effect of internal conversion on a long timescale always exists, even if the cavity is lossless. The cavity can ideally have almost the same free charge carrier generation rate as the bare system but cannot surpass it. An possible solution to the fast internal conversion is multiple-pulse pumping with an interval of 100fs to keep the system at the polariton-beneficial stage for "instantaneous" charge generation. 

Large cavity loss and fast internal conversion is a big challenge for polariton-mediated free charge carrier generation in the long timescale of picoseconds for the single-pulse pumping experiments.

\subsection{\label{sec:level3C}Effect of Collective Light-Matter Coupling Strength}

\begin{figure*}
    \includegraphics[width=0.9\textwidth]{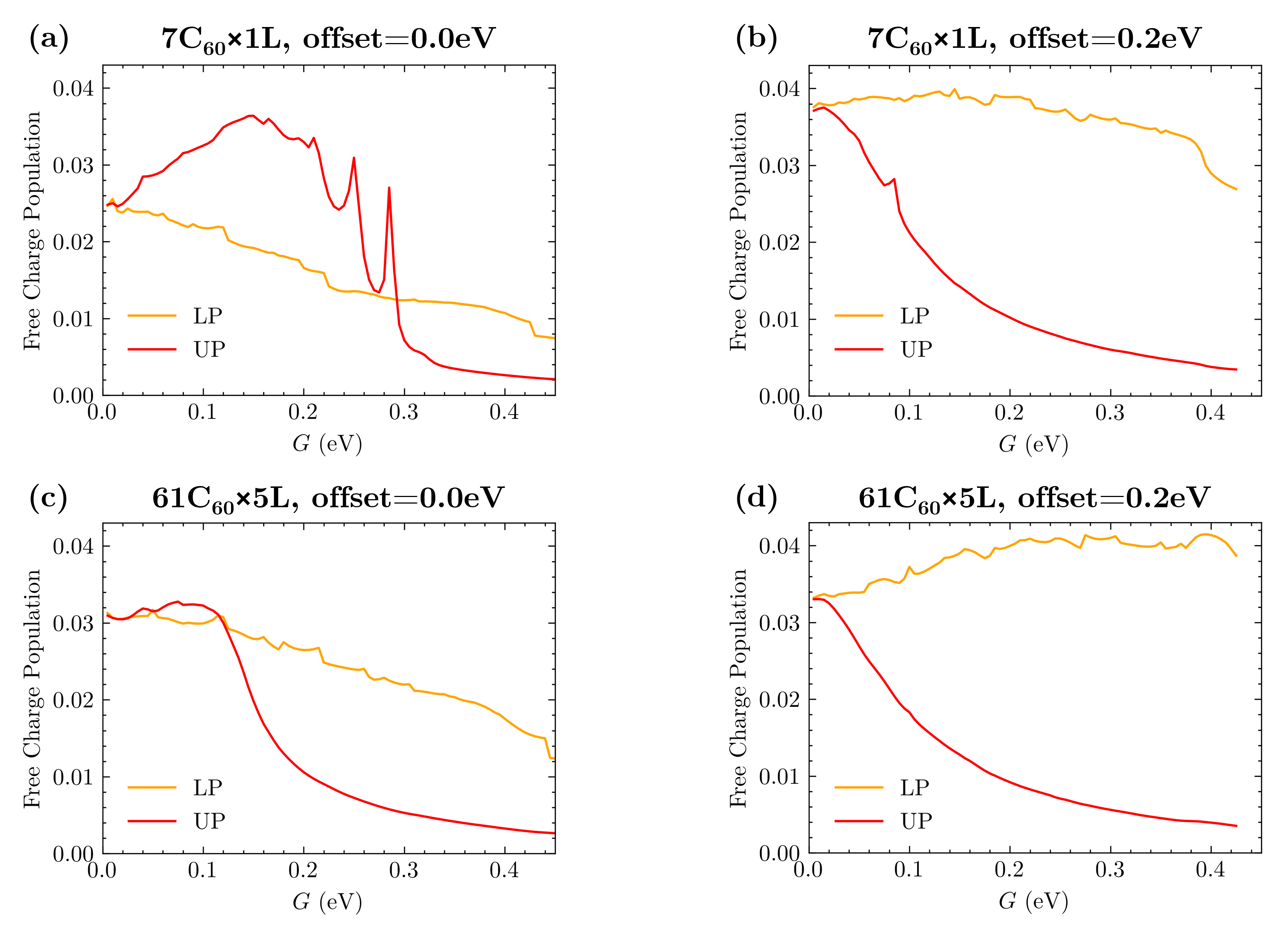}
    \caption{The free charge carrier population at time \SI{0.1}{\ps} when the cavity system is initially prepared at lower polariton (LP) or upper polariton (UP) with different collective light-matter coupling strength $G$. Four different conditions of fullerene packing type and interfacial energy offset are displayed: (a) 7C$_{60}\times1$L and $\Delta E_{\mathrm{offset}}=\SI{0.0}{\eV}$, (b) 7C$_{60}\times1$L and $\Delta E_{\mathrm{offset}}=\SI{0.2}{\eV}$, (c) 61C$_{60}\times5$L and $\Delta E_{\mathrm{offset}}=\SI{0.0}{\eV}$, (d) 61C$_{60}\times5$L and $\Delta E_{\mathrm{offset}}=\SI{0.2}{\eV}$.}
    \label{fig:CollectiveCoupling}
\end{figure*}

We next investigate the free charge carrier population generated within \SI{0.1}{\ps} versus the variations of the collective light-matter coupling $G$ to illustrate the "instantaneous" charge generator feature of polariton. $G$ can be experimentally modified by changing the effective coupling sites in each oligothiophene chain. The two mechanisms of the bare system charge separation\cite{tamura2013ultrafast} illustrated in section \ref{sec:level1} are also considered, including the electron delocalization in fullerenes and the vibronic charge transfer state induced by excess energy of interfacial XT-CS transition. The simulation results are summarized in Figure~\ref{fig:CollectiveCoupling}. The upper and lower polariton curves converge on the left, corresponding to the initial condition of a bare bright state.

In the same system studied above, i. e., packing type 7C$_{60}\times1$L and $\Delta E_{\mathrm{offset}}=\SI{0.2}{\eV}$ (Figure~\ref{fig:CollectiveCoupling}(a)), pumping upper polariton can be significantly optimized when $G=\SI{0.15}{\eV}$. Comparing the maximum value of upper polariton case (global maximum of the red curve in Figure~\ref{fig:CollectiveCoupling}(a)) with the bare system (left ends of the red curve in Figure~\ref{fig:CollectiveCoupling}(a)), we find that pumping upper polariton can enhance free charge population by 50\%. The delocalization feature of dark CS eigenstates (Figure~\ref{fig:eigen}(a)) explains the emergence of the optimal $G$. The relatively large transition dipoles and the large $\mathrm{CS}_N$ components of low energy dark CS eigenstates make them beneficial for polariton-dark state hybridization and free charge carrier generation. The high energy dark CS eigenstates, which mostly populates around $\mathrm{CS}_{N/2}$, is not as beneficial as the low energy dark CS eigenstates. Therefore, the upper polariton with excessive energy is not preferred because the corresponding dark CS state for second-hybridization is not optimal. The optimal coupling strength also relates to to the Marcus Inverted Region in which the charge transfer rate decreases with further increasing free energy. \cite{marcus1993electron} Moreover, the upper polariton-dark CS eigenstates hybridization effect is very obvious when the energy match at, for example, $G=\SI{0.04}{\eV}$ and $G=\SI{0.25}{\eV}$, where some peaks shows up with a local maximum of free charge carrier population.

As for the lower polariton, the free charge carrier population decreases as $G$ increases, while plenty of polariton-dark states hybridization peaks appear. For $G>\SI{0.3}{\eV}$, the lower polariton shows a larger free charge carrier population than that of the upper polariton, implying a larger photon loss when the upper polariton is not available for second-hybridization. 

When the Coulomb barrier is reduced to \SI{0.07}{\eV} by fullerene packing with 61C$_{60}\times5$L (Figure~\ref{fig:CollectiveCoupling}(c)), the upper polariton is optimized at $G=\SI{0.1}{\eV}$ with a flat peak. The population for lower polariton case is higher than that for upper polariton case when $G>\SI{0.12}{\eV}$, as a low Coulomb barrier is not suitable for upper polariton at high energy. When an interficial energy offset $\Delta E_{\mathrm{offset}}=\SI{0.2}{\eV}$ is added to previous two systems of different packing types (Figure~\ref{fig:CollectiveCoupling}(b)(d)), the free charge carrier generation is more efficient in bare system whereas in the cavity system upper polariton merely functions as a photon loss channel. The lower polariton can work as a bridge to connect the XT and CS manifolds in a low energy region due to their energy matching, increasing free charge carrier population. For packing type 61C$_{60}\times5$L with $\Delta E_{\mathrm{offset}}=\SI{0.2}{\eV}$ (Figure~\ref{fig:CollectiveCoupling}(d)), the bare system is in the Marcus Inverted Region, while lower polariton with lower energy shows better energy matching with the Coulomb barrier height, significantly enhancing free charge carrier generation. The result of cavity enhanced charge transfer in the Marcus Inverted Region is consistent with previous work\cite{semenov2019electron}. The eigenstates calculation and time evolution of density matrix for Figure~\ref{fig:CollectiveCoupling}(b-d) are listed in Supplementary Material S2.4.

All these results and discussion are based on H-type oligothiophene aggregate, while for the J-aggregate where the bright state lies on the lower edge, the lower polaritons can barely hybridize with any dark states with CS components and only upper polariton can enhance free charge carrier generation, as shown in Supplementary Material S2.3.

\section{\label{sec:level4}Conclusions}

We study a model oligothiophene:fullerene system to elucidate the effect of polariton mediated free charge generation on an ultrafast timescale. Based on a cavity QED Hamiltonian, an open quantum system method combining Redfield and Lindblad Theory is applied to study the quantum dynamics processes including vibration induced internal conversion and dephasing, cavity loss and electron-hole recombination. This general method can be further applied to extract features of the donor-acceptor photophysics in such molecular assemblies.

Our simulation anticipates that compared to the bright state of a bare H-aggregate donor system, the polariton has the potential to enhance free charge generation rate in the timescale of several hundred femtoseconds, owing to the following two mechanisms. First, the polariton with tailored energy can optimize the contradictory relation between the Coulomb attraction barrier and vibronic charge transfer state.\cite{tamura2013ultrafast, tamura2013potential, huix2015concurrent} More specifically, the upper polariton can help the system to overcome the high Coulomb attraction barrier, while the lower polariton functions better if the bare system is in the Marcus Inverted Region. Second, we propose a novel 
mechanism that a second-hybridization between polariton and dark states occurs under the condition of energy matching and non-zero dark state transition dipole. This second-hybridization can create perturbed polariton states that inherit the optical activity and transition strength from the polaritons and large CS components from certain dark states, which can work as instantaneous free charge carrier generators after photoexcitation with high efficiency within hundreds of femtoseconds, up to 50\% enhancement comparing to the bare system.

However, our simulation reveals that fast internal conversion and large cavity loss can suppress free charge carrier generation on a long timescale of picoseconds. Internal conversion eliminates the large CS components feature of the initial perturbed polaritons, and cavity loss introduces an additional dissipation channel. Thus, we explain the experimental results\cite{delpo2021polariton} that polariton shows negative effect, and further proposed that increasing cavity Q-factor or multiple-pulse pumping might solve this drawback. 

In totality, we theoretically show that despite their short lifetime, polaritons can still work as powerful tools to optimize the excitonic character on a short timescale without modifying the donor-acceptor materials' intrinsic properties (such as morphology optimization \cite{lee2018recent}). The free charge carrier enhancement is partly explained by the second-hybridization mechanism proposed herein. This novel insight here could be incorporated in designing the geometry and architecture of photovoltaics to achieve better performance.

\section*{Supplementary Material}
See Supplementary Material for technical details of exciton Hamiltonian, Bloch-Redfield Theory simulation, perturbation theory and supplementary results of simulated eigenstates and dynamics for systems with different parameters.

\begin{acknowledgments}
This research is funded by the U.S. Department of Energy, Office of Science, Office of Basic Energy Sciences Solar Photochemistry program, under Award Number DE-SC0015429. We acknowledge Abraham Nitzan, Adam Reinhold and Naser G. Mahfouz for the valuable discussion and reviewing and Alfy Benny for the assistance with Blender.

\end{acknowledgments}

\section*{AUTHOR DECLARATIONS}
\subsection*{Conflict of Interest}
The authors have no conflicts to disclose.

\section*{Data Availability Statement}
The data that support the findings of this study are available
from the corresponding author upon reasonable request.

\bibliography{aipsamp}

\onecolumngrid
\newpage 

\setcounter{section}{0}
\setcounter{equation}{0}
\setcounter{figure}{0}
\renewcommand\thesection{S\arabic{section}}
\renewcommand \thefigure {S\arabic{figure}}

\begin{center}
{\Large Supplementary Material \\ 
\vspace{0.2cm}
Polariton Enhanced Free Charge Carrier Generation in Donor–Acceptor Cavity Systems by a Second-Hybridization Mechanism
}
\end{center}

\section{Theoretical Details}
\subsection{Simplification of 2D Frankel Exciton Hamiltonian}
Here we write the 2D Oligothiophene Aggregate in real space 
\begin{equation}
H_{\mathrm{OT}}=\sum_{x,y}^{N_x, N_y}{\epsilon |\mathrm{XT}_{x,y}\rangle \langle \mathrm{XT}_{x,y}|}+\sum_{x,y}^{N_x, N_y}{\left( J_{\mathrm{XT}}|\mathrm{XT}_{x,y}\rangle \langle \mathrm{XT}_{x+1,y}|+J_y|\mathrm{XT}_{x,y}\rangle \langle \mathrm{XT}_{x,y+1}|+\mathrm{h}.\mathrm{c}. \right)}
\end{equation}
Assuming that there is periodic boundary condition for y-direction, one can turn the y-direction into momentum space $|\mathrm{XT}_{x,k}\rangle =\frac{1}{\sqrt{N_y}}\sum_y{e^{iky}|\mathrm{XT}_{x,y}\rangle}$, where $k=0,\frac{2\pi}{N_y},...,\frac{2\pi \left( N_y-1 \right)}{N_y}$. So the the Hamiltonian in x-real y-momentum space reads
\begin{equation}
H_{\mathrm{OT}}=\sum_k{\left( \sum_{x=1}^{N_x}{\left( \left( \epsilon +2J_y\cos k \right) |\mathrm{XT}_{x,k}\rangle \langle \mathrm{XT}_{x,k}|+J_x\left( |\mathrm{XT}_{x+1,k}\rangle \langle \mathrm{XT}_{x,k}|+\mathrm{h}.\mathrm{c}. \right) \right)} \right)}
\end{equation}
It is a block-diagonalized Hamiltonian with each block a 1D system. Choosing the bright block ($k=0$ so that the transition dipole is non-zero) to remove the momentum label will rebuild the Frankel Exciton Hamiltonian shown in the main text.

\subsection{Vibronic Coupling Hamiltonian}
Total Hamiltonian contains all both electronic and vibrational modes can be written as 
\begin{equation}
H_{\mathrm{total}}=H_{\mathrm{e}}+H_{\mathrm{v}}+H_{\mathrm{I}}
\end{equation}
, where the vibronic Hamiltonian and electronic-vibrational Interation are
\begin{equation}
H_{\mathrm{v}}=\sum_{l=1}^{N_{\mathrm{F}}}{\hbar \omega _{l}^{\mathrm{F}}\left( \left( b_{l}^{\mathrm{F}} \right) ^{\dagger}b_{l}^{\mathrm{F}}+\frac{1}{2} \right)}+\sum_{n=1}^{\max \left\{ N_{\mathrm{XT}},N_{\mathrm{CS}} \right\}}{\sum_{l=1}^{N_{\mathrm{OT}}}{\hbar \omega _{l}^{\mathrm{OT}}\left( \left( b_{n,l}^{\mathrm{OT}} \right) ^{\dagger}b_{n,l}^{\mathrm{OT}}+\frac{1}{2} \right)}}
\end{equation}
\begin{equation}
H_{\mathrm{I}}=\sum_{n=1}^{N_{\mathrm{XT}}}{B_{n}^{\mathrm{XT}}}|\mathrm{XT}_n\rangle \langle \mathrm{XT}_n|+\sum_{n=1}^{N_{\mathrm{CS}}}{B_{n}^{\mathrm{CS}}}|\mathrm{CS}_n\rangle \langle \mathrm{CS}_n|
\end{equation}
, where
\begin{equation}
B_{n}^{\mathrm{XT}}=\sum_{l=1}^{N_{\mathrm{OT}}}{\frac{c_{\mathrm{XT},l}^{\mathrm{OT}}}{\sqrt{2}}\left( \left( b_{n,l}^{\mathrm{OT}} \right) ^{\dagger}+b_{n,l}^{\mathrm{OT}} \right)}
\end{equation}
\begin{equation}
B_{n}^{\mathrm{CS}}=\sum_{l=1}^{N_{\mathrm{F}}}{\frac{c_{\mathrm{CS},l}^{\mathrm{F}}}{\sqrt{2}}\left( \left( b_{l}^{\mathrm{F}} \right) ^{\dagger}+b_{l}^{\mathrm{F}} \right)}+\sum_{l=1}^{N_{\mathrm{OT}}}{\frac{c_{\mathrm{CS},l}^{\mathrm{OT}}}{\sqrt{2}}\left( \left( b_{n,l}^{\mathrm{OT}} \right) ^{\dagger}+b_{n,l}^{\mathrm{OT}} \right)}
\end{equation}
The data of vibrational modes and coupling strength comes from Ref\citenum{huix2015concurrent} as shown in the table below. The unit is eV for all the data.

\begin{table}[]
\begin{tabular}{|c|c|c|c|c|c|}
\hline
Mode Index & $\omega _{l}^{\mathrm{F}}$ & $\omega _{l}^{\mathrm{OT}}$ & $c_{\mathrm{CS},l}^{\mathrm{F}}$ & $c_{\mathrm{CS},l}^{\mathrm{OT}}$ & $c_{\mathrm{XT},l}^{\mathrm{OT}}$ \\\hline
1          & 0.200025    & 0.401283     & 0.063988   & 0.009923      & 0.005707      \\
2          & 0.184269    & 0.397773     & 0.092915   & -0.00011      & 0.004131      \\
3          & 0.177853    & 0.182714     & -0.05697   & -0.09595      & -0.18344      \\
4          & 0.14111     & 0.178531     & -0.02476   & 0.081555      & 0.066305      \\
5          & 0.093952    & 0.13455      & 0.039635   & -0.05677      & -0.04654      \\
6          & 0.079933    & 0.111848     & -0.01928   & 0.016518      & 0.051747      \\
7          & 0.055892    & 0.042621     & -0.03356   & -0.01525      & -0.02858      \\
8          & 0.033264    & 0.018316     & 0.013944   & -0.01741      & -0.01099      \\ \hline
\end{tabular}
\end{table}
It is assumed that the Oligothiophene (OT) electronic-vibrational couplings are identical for all fragments. All the vibrational modes are independent to each other and satisfied the canonical commutation relation. Relabel each single interaction terms of the interaction Hamiltonian as $\mathrm{XT}_n\rightarrow N_{\mathrm{XT}}-n+1$ and $\mathrm{CS}_n\rightarrow N_{\mathrm{XT}}+n$ and we could also assume that $N_{\mathrm{XT}}=N_{\mathrm{OT}}=N$. Thus, we rewrite interaction Hamiltonian in an explicit way:
\begin{equation}
H_{\mathrm{I}}=\sum_{i=1}^N{B_i}S_i
\end{equation}

\subsection{Bloch-Redfield Theory for Vibronic Couplings}
Bloch-Redfield Theory can be applied to deal with the electronic-vibrational couplings by treating the electronic states as the system and vibrational states as the bath, namely a large thermal reservoir $\rho _{\mathrm{v}}=\frac{e^{H_{\mathrm{v}}/k_{\mathrm{B}}T}}{\mathrm{Tr}\left( e^{H_{\mathrm{v}}/k_{\mathrm{B}}T} \right)}$, when Born Approximation and Markovian Approximation hold under the condition of weak electronic-vibrational couplings. The Quantum Master Equation in Schrödinger Picture reads
\begin{equation}
\begin{aligned}
\frac{i}{\hbar}\left[ H_{\mathrm{e}},\rho _{\mathrm{e}}\left( t \right) \right] +\frac{d\rho _{\mathrm{e}}\left( t \right)}{dt}
={}&
-\frac{1}{\hbar ^2}\sum_{ij}{I_{ij}\left( -\omega _c+\omega _d \right) \langle a|S_i|b\rangle \langle c|S_j|d\rangle \left[ |a\rangle \langle b|,|c\rangle \langle d|\rho _{\mathrm{e}}\left( t \right) \right]}
{}\\&
-\frac{1}{\hbar ^2}\sum_{ij}{\sum_{abcd}{I_{ij}^{*}\left( \omega _c-\omega _d \right)}}\langle a|S_i|b\rangle \langle c|S_j|d\rangle \left[ \rho _{\mathrm{e}}\left( t \right) |c\rangle \langle d|,|a\rangle \langle b| \right] 
\end{aligned}
\end{equation}
Here $i,j$ are the labels for each electronic-vibrational coupling terms and $a,b,c,d$ are the labels for the eigenstates of electronic Hamiltonian $H_{\mathrm{e}}$. Notice that secular approximation is not considered. The integral is defined as 
\begin{equation}
I_{ij}\left( \omega \right) =\int_0^{\infty}{\mathrm{Tr}\left( \tilde{B}_i\left( \tau \right) \tilde{B}_j\left( 0 \right) \rho _{\mathrm{v}} \right) e^{i\omega \tau}d\tau}
\end{equation}
The tildes represent Dirac Picture. By expanding the $\tilde{B}_i\left( \tau \right)$ and $\tilde{B}_j\left( 0 \right)$, and ignoring Lamb Shift, the integral can be expressed explicitly as 
\begin{equation}
I_{ij}\left( \omega \right)=\pi J_{ij}\left( -\omega \right) \bar{n}\left( -\omega \right) +\pi J_{ij}\left( \omega \right) \left( \bar{n}\left( \omega \right) +1 \right) 
\end{equation}
, where $\bar{n}\left( \omega \right) =\left( e^{\hbar \omega /k_{\mathrm{B}}T}-1 \right)^{-1}$ is Bosonic Distribution and $J_{ij}\left( \omega \right)$ depends on the three cases of $i,j$ combinations:
\begin{equation}
J_{ij}\left( \omega \right) =
\begin{cases}
	\delta _{ij}J_{\mathrm{XT},\mathrm{XT}}^{\mathrm{OT}}\left( \omega \right) ,                      &\mathrm{when} \ i\rightarrow \mathrm{XT}_n, j\rightarrow \mathrm{XT}_m\\
	J_{\mathrm{CS},\mathrm{CS}}^{\mathrm{F}}\left( \omega \right) +\delta _{ij}J_{\mathrm{CS},\mathrm{CS}}^{\mathrm{OT}}\left( \omega \right) ,       &\mathrm{when}\ i\rightarrow \mathrm{CS}_n, j\rightarrow \mathrm{CS}_m\\
	\delta _{N_{\mathrm{XT}}+1-i,j-N_{\mathrm{XT}}}J_{\mathrm{XT},\mathrm{CS}}^{\mathrm{OT}}\left( \omega \right) ,        &\mathrm{when}\ i\rightarrow \mathrm{XT}_n, j\rightarrow \mathrm{CS}_m\\
\end{cases}
\end{equation}
, where
\begin{equation}
\begin{aligned}
J_{\mu ,\nu}^{X}\left( \omega \right) =\sum_{l=1}^{N_X}{\frac{1}{2}c_{\mu ,l}^{X}c_{\nu ,l}^{X}\delta \left( \omega -\omega _{l}^{X} \right)}
\end{aligned}
\end{equation}
with $X=\mathrm{OT},\mathrm{F}$ and $\mu ,\nu =\mathrm{XT},\mathrm{CS}$. Numerically, Dirac Delta Function can be approximately replaced by Lorentzian Function $\delta \left( \omega -\omega _0 \right) \approx \frac{1}{\pi \gamma}\left[ 1+\left( \frac{\omega -\omega _0}{\gamma} \right) ^2 \right] ^{-1}$ with $\gamma = 0.01\mathrm{eV}$.

\subsection{Perturbation Theory}
Rewrite the Electronic Hamiltonian in the bare eigen-basis
\begin{equation}
H_{\mathrm{e}}=\sum_{\alpha}{E_{\alpha}}|\alpha \rangle \langle \alpha |+\hbar \omega _{\mathrm{c}}|1\rangle \langle 1|+\sum_{\alpha}{\left( g_{\alpha}^{'}|\alpha \rangle \langle 1|+\left( g_{\alpha}^{'} \right) ^*|1\rangle \langle \alpha | \right)}
\end{equation}
, where $g_{\alpha}^{'}=\vec{\varepsilon}\cdot \langle \alpha |\vec{\mu}|\mathrm{vac}\rangle$ is the dipole field interaction. Based on the transition dipole of each bare eigenstates, we can define the one and only one bright state for the bare system as $g_{\mathrm{B}}^{'}=\max \left\{ g_{\alpha}^{'} \right\} , g_{\mathrm{B}}^{'}\gg g_{\alpha \ne \mathrm{B}}^{'}$, which is valid for the model we study and is the origin of 2 polariton states. The eigenstates other than the bright state are the dark states $|d\ne \mathrm{B}\rangle $. Next we adjust the cavity mode resonant with the bare bright state $\hbar \omega _{\mathrm{c}}=E_{\mathrm{B}}$, and define Upper Polariton $|\mathrm{UP}\rangle =\frac{1}{\sqrt{2}}\left( |1\rangle +e^{i\mathrm{arg}\left( g_{\mathrm{B}}^{'} \right)}|\mathrm{B}\rangle \right) $ and Lower Polariton $|\mathrm{LP}\rangle =\frac{1}{\sqrt{2}}\left( |1\rangle -e^{i\mathrm{arg}\left( g_{\mathrm{B}}^{'} \right)}|\mathrm{B}\rangle \right) $. Now we can split the Electronic Hamiltonian $H_\mathrm{e}$ into the unperturbed Hamiltonian $H_0$ (energy terms of polaritons and dark states) and the Perturbation $H^{'}$ (weak dark-light coupling) in the basis consisting of polaritons and dark states:

\begin{equation}
H_0 =\left( E_{\mathrm{B}}+\left| g_{\mathrm{B}}^{'} \right| \right) |\mathrm{UP}\rangle \langle \mathrm{UP}|+\left( E_{\mathrm{B}}-\left| g_{\mathrm{B}}^{'} \right| \right) |\mathrm{LP}\rangle \langle \mathrm{LP}|+\sum_d{E_d}|d\rangle \langle d|
\end{equation}
\begin{equation}
H^{'} =\sum_d{\left( \frac{g_{d}^{'}}{\sqrt{2}}|d\rangle \langle \mathrm{LP}|+\frac{g_{d}^{'}}{\sqrt{2}}|d\rangle \langle \mathrm{UP}|+\frac{\left( g_{d}^{'} \right) ^*}{\sqrt{2}}|\mathrm{LP}\rangle \langle d|+\frac{\left( g_{d}^{'} \right) ^*}{\sqrt{2}}|\mathrm{UP}\rangle \langle d| \right)}
\end{equation}
Eigenstate up to first-order perturbation correction is
\begin{equation}
\begin{aligned}
|\mathrm{LP}^{\left( 1 \right)}\rangle &=|\mathrm{LP}\rangle +\sum_d{\frac{{{g_{d}^{'}}/{\sqrt{2}}}}{E_{\mathrm{B}}-\left| g_{\mathrm{B}}^{'} \right|-E_d}|d\rangle}
\\
|\mathrm{UP}^{\left( 1 \right)}\rangle &=|\mathrm{UP}\rangle +\sum_d{\frac{{{g_{d}^{'}}/{\sqrt{2}}}}{E_{\mathrm{B}}+\left| g_{\mathrm{B}}^{'} \right|-E_d}|d\rangle}
\\
|d^{\left( 1 \right)}\rangle &=|d\rangle +\frac{{{\left( g_{d}^{'} \right) ^*}/{\sqrt{2}}}}{E_d-\left( E_{\mathrm{B}}+\left| g_{\mathrm{B}}^{'} \right| \right)}|\mathrm{UP}\rangle +\frac{{{\left( g_{d}^{'} \right) ^*}/{\sqrt{2}}}}{E_d-\left( E_{\mathrm{B}}-\left| g_{\mathrm{B}}^{'} \right| \right)}|\mathrm{LP}\rangle 
\end{aligned}
\end{equation}
Here we can see a mixing between unperturbed polaritons and dark states.

\newpage
\section{Supplementary Results}
\subsection{Effect of Cavity Loss}
Introducing cavity leads to significant dissipation, especially for the upper polariton case. If the lossless cavity is applied, the rate curve of cavity system will have the same tendency as the rate of the lossy cavity in short time, but it will converge to the rate curve of the bare systems on large time due to extreme small photon components. (Actually the rates for cavity cases are slightly smaller than the bare cases because of photon population contribution)

\begin{figure}[htbp]
    \includegraphics[width=\textwidth]{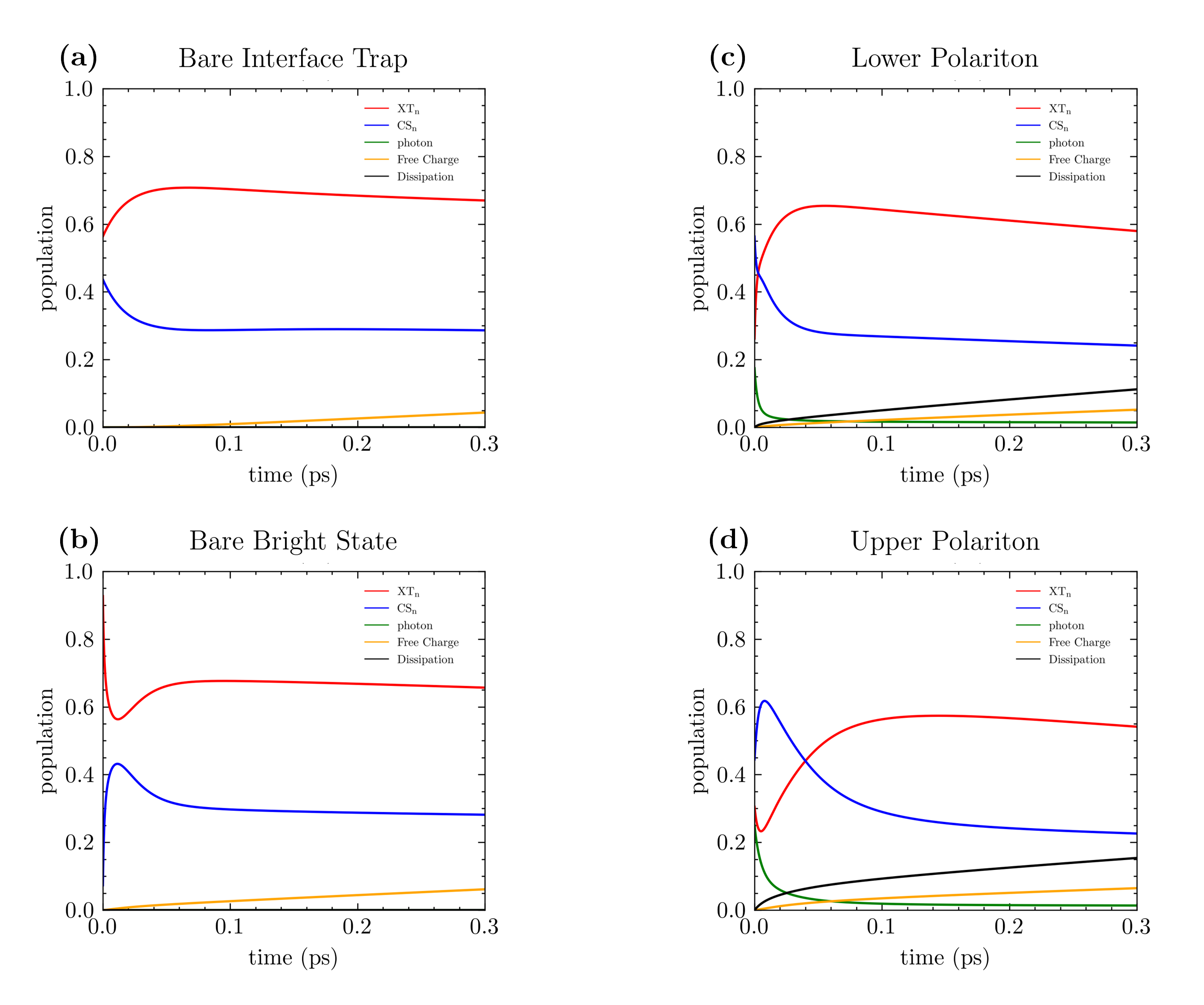}
    \caption{Extracting population dynamics of each catalogues of species from Figure 3 for the dynamics of the four cases along with the dynamics of the free charge carrier and dissipation.}
    \label{fig:s1}
\end{figure}

\newpage
\begin{figure}[htbp]
    \includegraphics[width=0.6\textwidth]{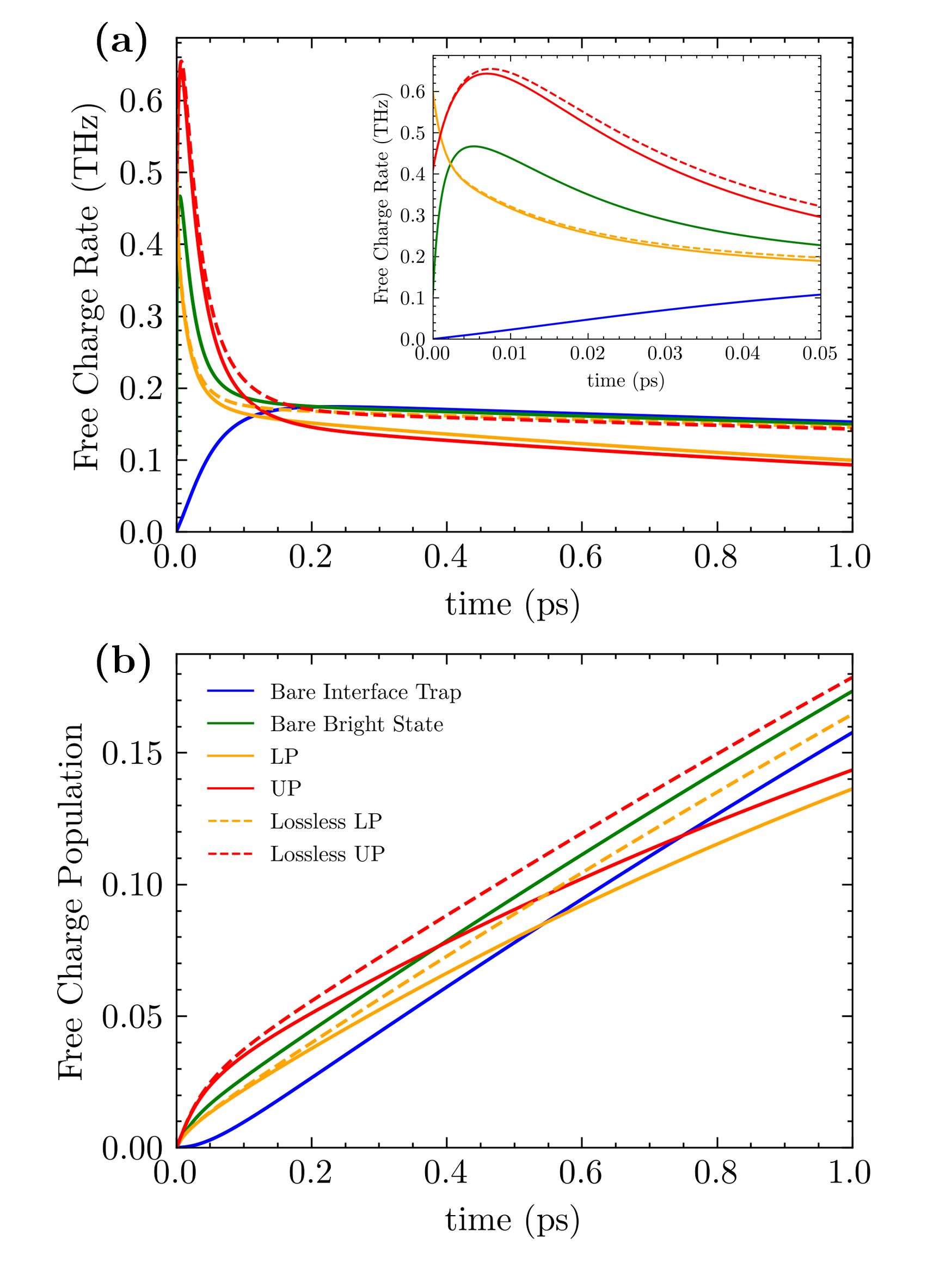}
    \caption{A comparison between normal system and lossless system.(a) Free charge carrier generation rate at each time step up to 1ps. The inserted figure is the zoomed in area of the first 0.05ps. (b) Free charge carrier population dynamics by integration for the corresponding rates in (a).}
    \label{fig:s2}
\end{figure}

\newpage
\subsection{Effect of Inhomogeneous Broadening}
The homogeneous broadening is taken to be very small so that each absorption peak is distinguishable. Different upper polaritons/lower polaritons would converge respectively.

\begin{figure}[htbp]
    \includegraphics[width=0.92\textwidth]{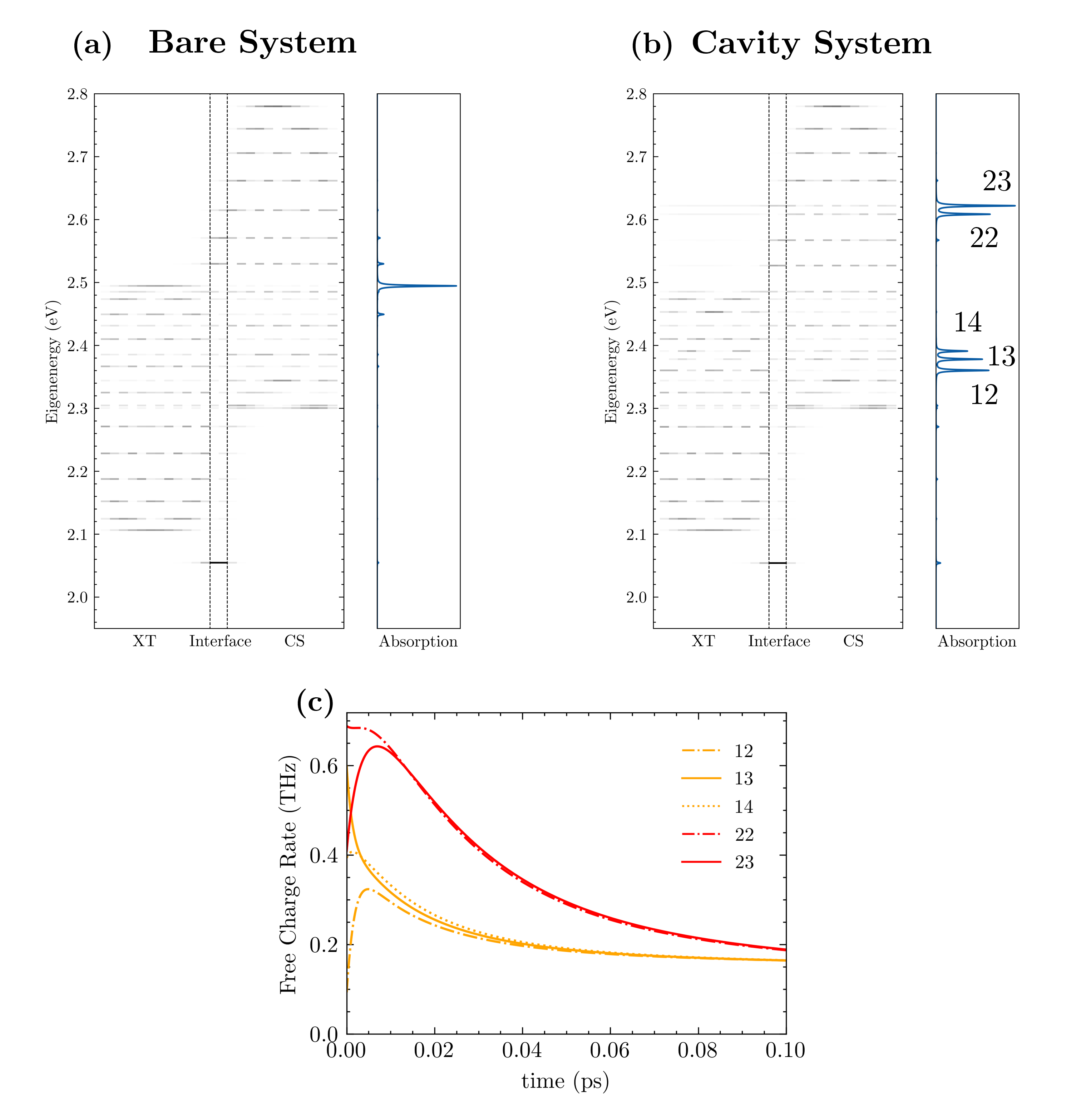}
    \caption{(a)(b) Remake of Figure 2 to better display inhomogeneous broadening. (c) initial condition for different upper polaritons (red) and lower polaritons (orange).}
    \label{fig:s3}
\end{figure}

\newpage
\subsection{Effect of J-Aggregate}

\begin{figure}[htbp]
    \includegraphics[width=0.92\textwidth]{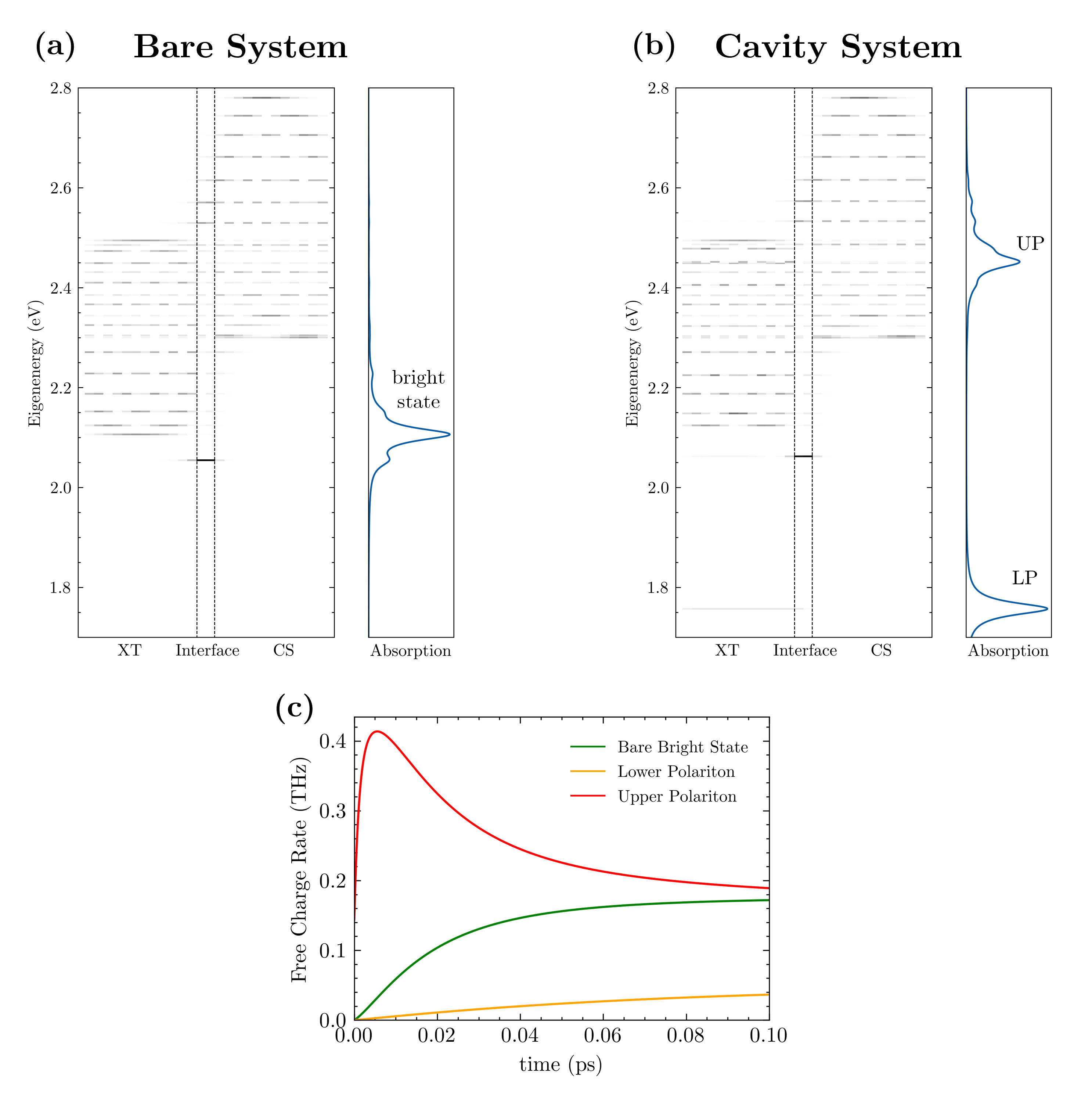}
    \caption{Electronic excited eigenstate of a J-aggregate system with fullerene packing type 7C$_{60}\times1$L and $\Delta E_{\mathrm{offset}}=0\mathrm{eV}$ for (a) Bare System and (b) Cavity System ($\omega_\mathrm{c}=2.495\mathrm{eV}$ and $G=0.35\mathrm{eV}$). (c) The free charge carrier generation rate at each time step for different initial condition in a J-aggregate system.}
    \label{fig:s4}
\end{figure}

\newpage
\subsection{Eigenvector and Dynamics Analysis}

\begin{figure}[htbp]
    \includegraphics[width=0.92\textwidth]{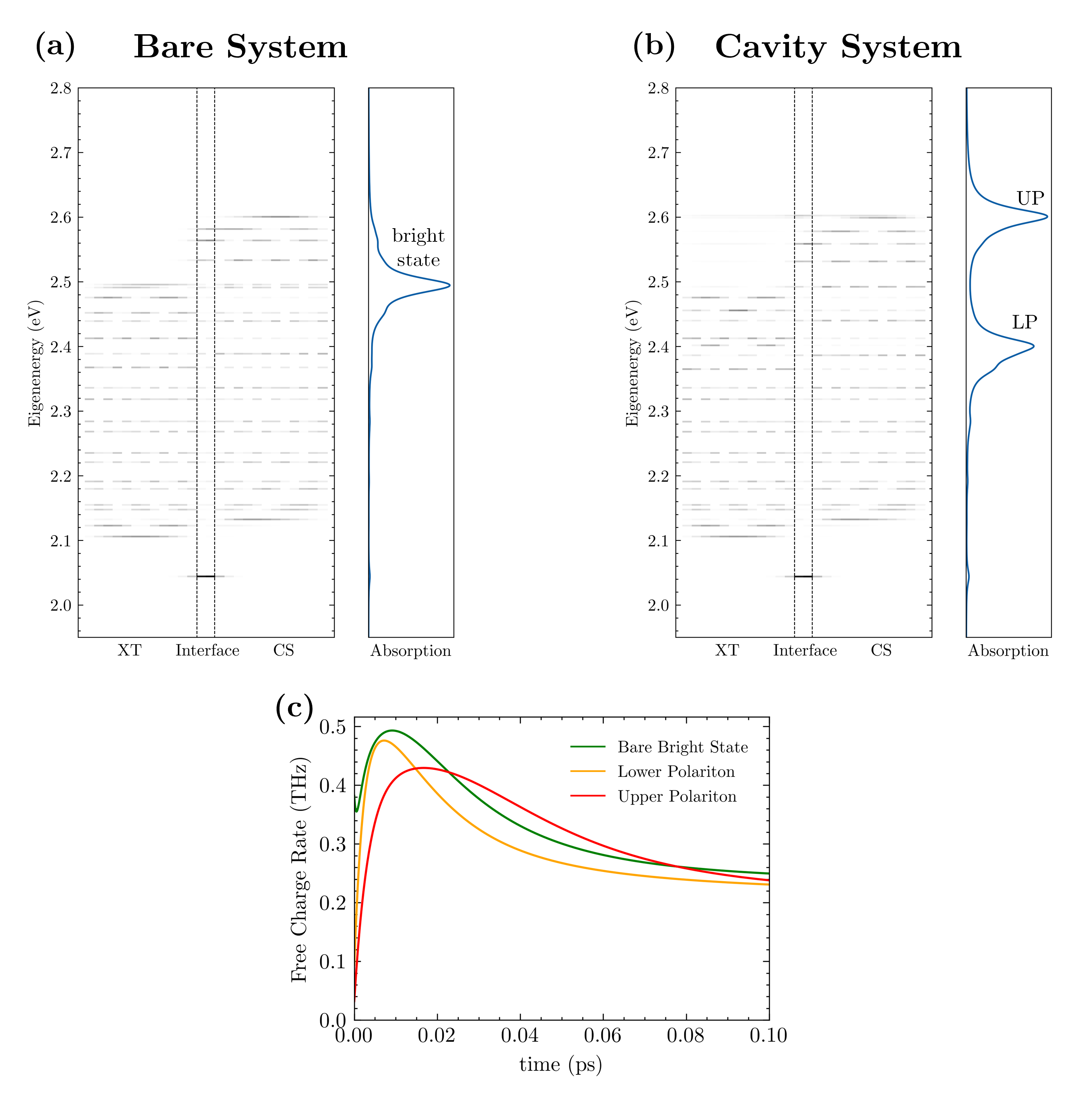}
    \caption{Electronic excited eigenstate of a H-aggregate system with fullerene packing type 61C$_{60}\times5$L and $\Delta E_{\mathrm{offset}}=0\mathrm{eV}$ for (a) Bare System and (b) Cavity System ($\omega_\mathrm{c}=2.495\mathrm{eV}$ and $G=0.10\mathrm{eV}$). (c) The free charge carrier generation rate at each time step for different initial condition in a J-aggregate system.}
    \label{fig:s5}
\end{figure}

\begin{figure}[htbp]
    \includegraphics[width=0.92\textwidth]{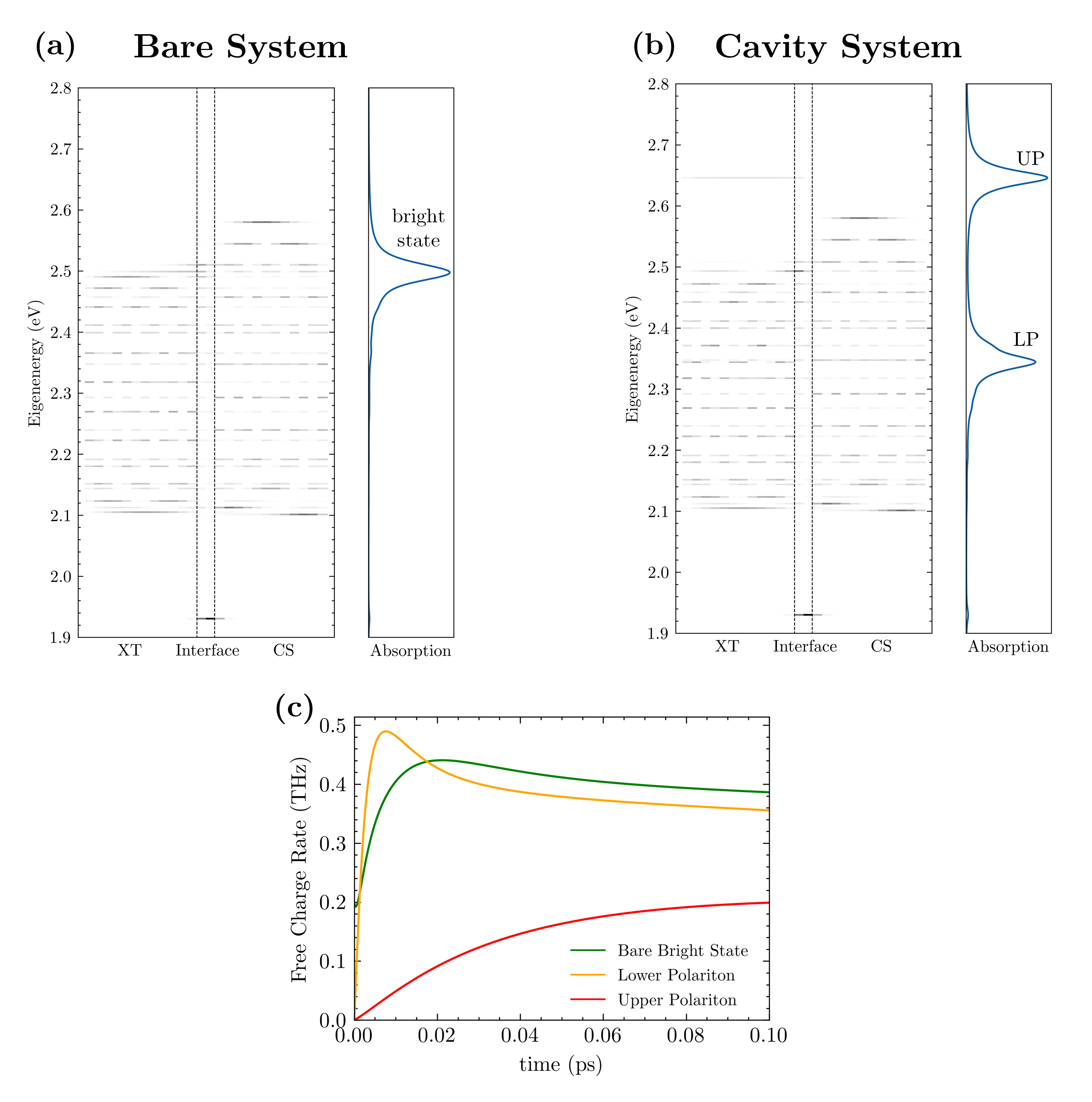}
    \caption{Electronic excited eigenstate of a H-aggregate system with fullerene packing type 7C$_{60}\times1$L and $\Delta E_{\mathrm{offset}}=0.2\mathrm{eV}$ for (a) Bare System and (b) Cavity System ($\omega_\mathrm{c}=2.495\mathrm{eV}$ and $G=0.15\mathrm{eV}$). (c) The free charge carrier generation rate at each time step for different initial condition in a J-aggregate system.}
    \label{fig:s6}
\end{figure}

\begin{figure}[htbp]
    \includegraphics[width=0.92\textwidth]{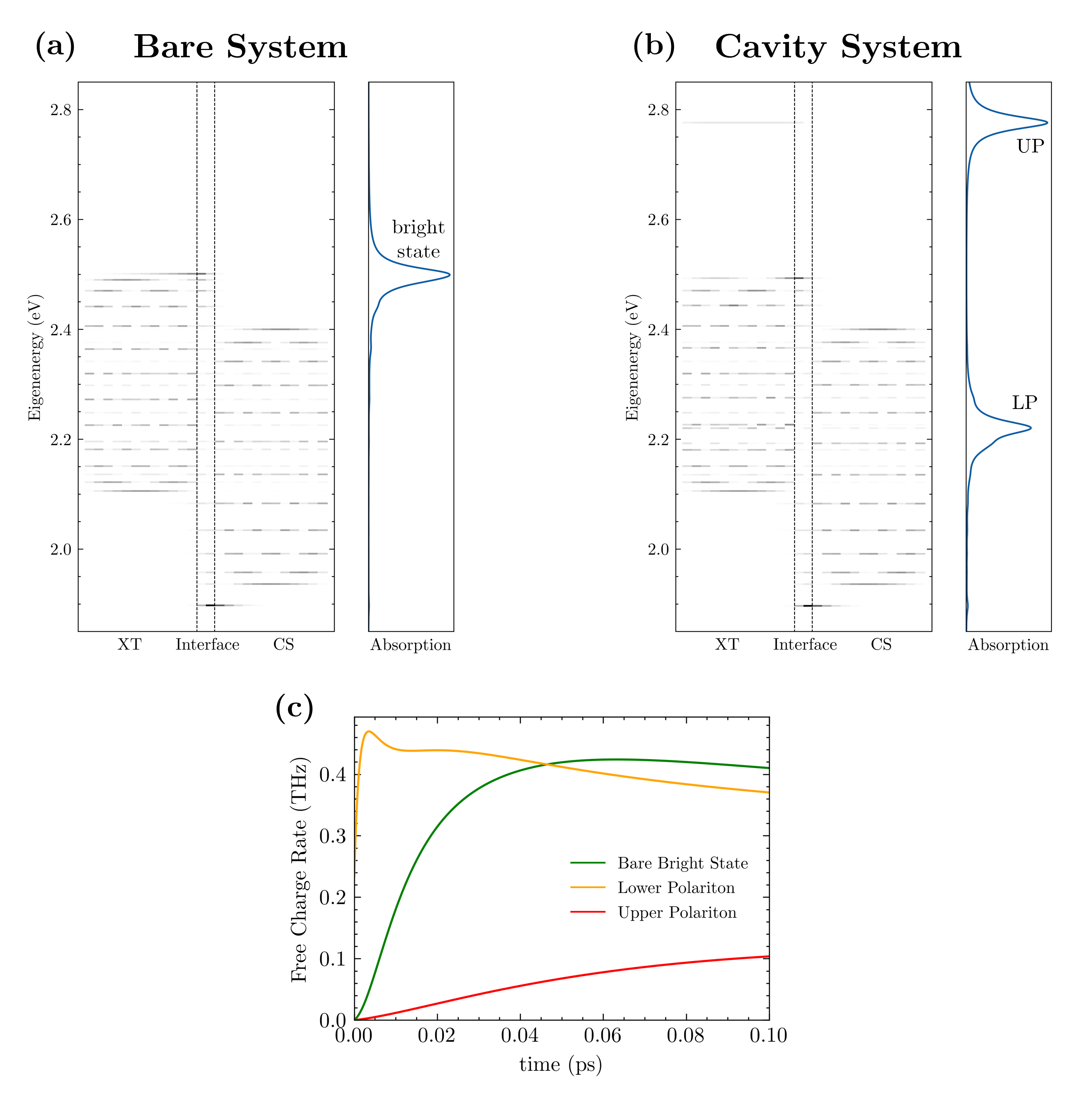}
    \caption{Electronic excited eigenstate of a H-aggregate system with fullerene packing type 61C$_{60}\times7$L and $\Delta E_{\mathrm{offset}}=0.2\mathrm{eV}$ for (a) Bare System and (b) Cavity System ($\omega_\mathrm{c}=2.495\mathrm{eV}$ and $G=0.28\mathrm{eV}$). (c) The free charge carrier generation rate at each time step for different initial condition in a J-aggregate system.}
    \label{fig:s7}
\end{figure}

\end{document}